\documentclass[a4paper,10pt]{article}
\usepackage{amsmath,amssymb}
\newcommand{\beq}{\begin{equation}}
\newcommand{\eeq}{\end{equation}}
\newcommand{\beqnon}{\begin{displaymath}}
\newcommand{\eeqnon}{\end{displaymath}}
\newcommand{\be}{\begin{equation}}
\newcommand{\ee}{\end{equation}}

\newcommand{\bx}{\bar{x}}
\newcommand{\bz}{\bar{z}}
\newcommand{\by}{\bar{y}}

\newcommand{\pd}{\partial}
\newcommand{\beqa}{\begin{eqnarray}}
\newcommand{\eeqa}{\end{eqnarray}}
\newcommand{\hj}{\hat{\jmath}}

\begin{document}
\begin{flushright}
{\normalsize ROM2F/36/03}
\end{flushright}

\begin{center} {\Large
 On Hpp-wave/CFT$_2$ Holography} \footnote{Based on the talk
given at the `Tenth Marcel Grossmann Meeting', 20-26 July 2003,
CBPF-ICRA, Rio de Janeiro, Brazil.} \end{center} \begin{center}
{\large Oswaldo Zapata}\end{center}\begin{center}
 \emph {Dipartimento di Fisica, Universit\`a di Roma ``Tor
Vergata''}\\ \emph{INFN, Sezione di Roma ``Tor Vergata''}\\
\emph{Via della Ricerca Scientifica 1, I-00133, Rome, Italy}
\end{center}

 \begin{abstract}
 \noindent We briefly review the $AdS_3/CFT_2$ correspondence and
the holographic issues that arise in the Penrose limit. Exploiting
current algebra techniques, developed by D'Appollonio and Kiritsis
for the closely related Nappi-Witten model, we obtain preliminary
results for bosonic string amplitudes in the
resulting Hpp-wave background and comment on how to extend them to the superstring.\\

\end{abstract}

The $AdS_3/CFT_2$ holographic correspondence relates superstring
theory on $AdS_3\times S^3\times \mathcal{M}$ to a two-dimensional
superconformal field theory defined as the non-linear
$\sigma$-model whose target space is the symmetric orbifold
${\mathcal M}^N/{\mathcal S}_N$ \cite{malda,agmoo,gks}. The four
dimensional compactification manifold, $\mathcal M$, is chosen to
be $T^4$ or $K3$.

$AdS_3\times S^3\times \mathcal{M}$  can be thought of as the near
horizon geometry of a D1-D5 brane configuration with non-vanishing
R-R 3-form flux. Since it is not fully known how to quantize the
superstring in the presence of generic R-R backgrounds\footnote{At
present, the Pure Spinor formalism is the most promising approach
to tackle this outstanding problem \cite{berk}.}, it is convenient
to consider the S-dual configuration which is supported by a NS-NS
3-form flux \cite{gks}. The metric of the resulting F1-NS5 bound
state is \beqnon ds^2=f_1^{-1}\,(\,
-dx_0^2+dx_1^2\,)+f_5\,(\,dr^2+r^2d\Omega_3^2\,)+ds^2_{{}_{\mathcal{M}}}\,
, \eeqnon \beqnon f_1=1+\frac{g_s^2\alpha'Q_1}{vr^2}\, ,\quad
\quad f_5=1+\frac{\alpha'Q_5}{r^2}\, , \eeqnon where $f_1, \, f_5$
are harmonic functions in the transverse space, $Q_1$, $Q_5$ are
the number of fundamental strings and 5-branes, respectively, and
$v$ is the volume of ${\mathcal{M}}$ in string units. $AdS_3\times
S^3 \times \mathcal{M}$ (in Poincar\'e coordinates) emerges in the
near horizon limit $r\to 0$ after the change of variables
$g_s^2=v\, Q_5/Q_1$ and $ r=R^2/u=Q_5\, \alpha
'/u$.\\

In principle, one can consider $AdS_3\times S^3$ supported by both
R-R and NS-NS 3-form fluxes, where the supergravity equations give
\beqnon R^2=\sqrt{(Q_5^{{}^{NS}})^2+g_s^2\,(Q_5^{{}^R})^2}\, .
\eeqnon The six-dimensional hybrid formalism of
Berkovits-Vafa-Witten (BVW) \cite{bvw} is suitable for
quantization in these backgrounds and
allows to compute some supersymmetric amplitudes \cite{dolan}.\\

Since the bosonic string provides useful insights into some basic
aspects of holography, we will first concentrate on this simpler
yet interesting case \cite{gks,mo}. The bosonic string on
$AdS_3\times S^3$ can be described by a Wess-Zumino-Novikov-Witten
(WZNW) model with left and right affine algebras
$\widehat{SL}(2,{\bf R})\times \widehat{SU}(2)$, which is a
completely solvable conformal field theory. Denoting by  $K^A(z)$
the generators of $\widehat{SL}(2,{\bf R})$ and those of
$\widehat{SU}(2)$ by $J^a(z)$, we can write the current algebras
as follows {\setlength\arraycolsep{2pt} \beqa \widehat{SL}(2,{\bf
R}): \qquad K^A(z)K^B(w) &\sim& \displaystyle\frac{k\,
\eta^{AB}}{2\,(z-w)^2} + i \,\epsilon^{ABC} \frac{ K^C(w)}{z-w}\,
, \nonumber\\\widehat{SU}(2):\qquad\quad J^a(z)J^b(w) &\sim&
\displaystyle \frac{k\,\delta^{ab}}{2\,(z-w)^2}+ i
\,\epsilon^{abc} \frac{ J^c(w)}{z-w}\, ,\nonumber \eeqa} \noindent
where the Cartan Killing metric for $\widehat{SL}(2,{\bf R})$ is
chosen to be $\eta^{AB}=\textrm{diag}\,(+,+,-)$. Here we are
considering the case $k_{SU(2)}+2 = k_{SL(2)} - 2 \gg 1$  and $c_{int}=20$.\\

 The problem of quantizing the
bosonic string in backgrounds containing an $AdS_3$ factor has a
long story \cite{petropoulos}. Only recently it has become clear
that the spectrum consists not only of short strings corresponding
to ``discrete'' representations with $1/2 <j < (k-1)/2$ and long
strings corresponding to ``continuous'' representations with $j =
1/2 + i s$, but also of spectral flowed representations
characterized by an integer $w$, to some extent similar to a
winding number \cite{mo}. String amplitudes with a low number of
insertions (two, three and four) can be computed via current
algebra methods or the Wakimoto free field representation
\cite{mo, teschner, hosom}. Extension to the superstring should
not present any major obstacle and we plan to address this issue
in the near future \cite{bdkz2}. Yet quantitative comparison with
boundary conformal field theory predictions is hampered by the
presence of moduli deformations of the non-linear $\sigma$-model
\cite{mo,seibwit} which would be lifted by turning on a R-R
background.

As mentioned above this is not fully under control, except
possibly for the Hpp-wave which is the local geometry seen by an
observer moving at the speed of light in $AdS_3\times S^3$. This
process of zooming-in around a null geodesic is called Penrose
limit \cite{penrose,blau}.

 In global coordinates,
$AdS_{3}\times S^{3}$ has the form \beqnon
  ds^{2} = R^{2} [
-(\cosh\rho)^{2} dt^{2} + d\rho^{2} + (\sinh\rho)^{2}
d\varphi_{1}^{2} + (\cos\theta)^{2} d\psi^{2} + d\theta^{2} +
(\sin\theta)^{2} d\varphi_{2}^{2}]\, , \eeqnon
  where $(t,\rho,\varphi_1)$
are the coordinates on $AdS_3$  and $(\theta,\psi,\varphi_2)$ are
those on $S^3$. In order to perform  the Penrose limit, we change
variables to \beqnon
  t ={u\over{2}} + { v \over  R^{2}}\, , \quad\qquad \psi =
\frac{u}{2} - { v \over  R^{2}}\, , \quad\qquad \rho = {r_{1}\over
R}\, , \quad\qquad \theta = {r_{2}\over R}\, , \eeqnon and take
the limit of infinitely large radius for both spaces. A useful
change of variables is
$r_i\exp(i\varphi_i)=\exp(iu/2)w_i$.\\

We note here that an unexpected $SU(2)_I$ symmetry comes from the
fact that the symmetry  of the 4-d transverse plane is broken by
the NS-NS 3-form flux $H_{+12} = H_{+34}=1$ according to $SO(4)\to
SU(2)_I\times U(1)_{J}$ \footnote{$H_{+12} = \mu_1 \neq
H_{+34}=\mu_2$ gives an exactly solvable CFT on the string
worldsheet, too. We will not discuss the general case here (see
\cite{bdkz1} ).} . Taking into account this extra symmetry, the
Hpp-wave metric takes the manifest $SU(2)_I$ invariant form
 \beqnon
  ds^{2} =
- 2 du dv + {i\over 4} du \sum_{\alpha=1}^{2}(
w^{\alpha}d\bar{w}_{\alpha} - \bar{w}_{\alpha}dw^{\alpha}) +
\sum_{\alpha=1}^{2} dw^{\alpha}d\bar{w}_{\alpha}\, .
 \eeqnon
 As usual we
can introduce momenta along the light-cone directions, $j\equiv
-i\,\pd_u$ and $p\equiv -i\,\pd_v$.\\

In line with what we said above we expect string theory in this
pp-wave background to be described by a special kind of WZNW
model. In particular, for the Penrose limit of $AdS_3\times S^3$,
the worldsheet CFT is a generalization of the Nappi-Witten (NW)
model, which in turn emerges from the Penrose limit of the near
horizon geometry of a stack of NS5-branes \cite{kda}. In our case
the Heisenberg algebra is six-dimensional ($\mathcal H_6$), while
for the standard NW model it is four-dimensional ($\mathcal H_4$).
From the current algebra point of view the Penrose limit is
carried out by contracting the currents of both CFTs as \beqnon
  K(z) = {i\over k}\,
[J^{3}(z) -K^{3}(z)]\, , \qquad J(z) = -i\, [J^{3}(z) +
K^{3}(z)]\, , \eeqnon and taking the limit $k\to \infty$. The
other generators are just scaled by the constant factor
$\sqrt{2/k}$. We obtain a Heisenberg algebra
 \beqnon
[P^+_{\alpha},P^{-\beta}]=-2i\delta_{\alpha}^{\beta}K\, ,
 \eeqnon
  \beqnon
[J\,,P^+_{\alpha}]=-iP^+_{\alpha}\, ,  \hspace{1.5cm}
[J\,,P^{-\alpha}]=+ iP^{-\alpha}\, . \eeqnon
 It should be stressed that current
contraction  (\`a la Saletan) is a general procedure that can be
applied to other models \cite{sfetsos}, {\it e.g.} to the
superstring in the already mentioned BVW approach
\nolinebreak\footnote{I thank N. Berkovits for pointing out this
possibility.}.

Exploiting  current algebra techniques developed for the NW model
by D'Appollonio and Kiritsis \cite{kda}, or the alternative free
field Wakimoto representation proposed by Cheung, Freidel and
Savvidy \cite{cfs}, one can derive explicit expressions for some
string amplitudes in the Hpp-wave background. Here we only report
some preliminary results. An extensive analysis will be
presented in a forthcoming paper \cite{bdkz1}. \\

First of all recall that $\mathcal H_6$ has three types of
representations, depending on the value of the light cone momentum
$p$. States can have $p\neq 0 $ (discrete representations) or
$p=0$ (continuous representations). The primary fields that we can
construct are of the general form $\Phi_{q}^{a}(z,\bz;x,\bx)$,
where $a$ specifies the type of representation, $a={\pm,0}$ for $p
\gtrless 0$ or $p=0$, respectively, $q$ stands for the momenta,
$q=(p,\hj\,)$ for $p\neq 0$ and $q=(s,\hj\,)$ for $p=0$, and $x$,
with the appropriate index ($x_{\alpha}$ for $p\geqslant 0$ and
$x^{\alpha}$ for $p<0$), are some auxiliary ``charge'' variables
that compactly encode all the states in a given representation.
Note that in this covariant approach we naturally include $p=0$
states,
inaccessible to the light cone quantization.\\
Using the charge variables, $\widehat{\mathcal H}_6$ can be
realized in terms of differential operators
 \beqnon
 p>0: \qquad P^+_{\alpha}=\sqrt2\,p\,x_{\alpha}, \qquad
 P^{-\,\alpha}=\sqrt2\,\pd^{\alpha}, \qquad
 J=i\,\left(\hj +x_{\alpha}\pd^{\alpha}\right),\quad K=ip\, ,
 \eeqnon
 \beqnon
 p<0: \qquad P^+_{\alpha}= \sqrt2\,\pd_{\alpha},\qquad
 P^{-\,\alpha}= \sqrt2\,p\,x^{\alpha}, \qquad
 J=i\,\left(\hj -x^{\alpha}\pd_{\alpha}\right), \quad K=-ip\, .
 \eeqnon
From now on we use the notation $p\equiv |p|$ for states with
$p<0$.

 With this realization of the algebra the two-point function
is easily obtained by imposing the Ward identities
 \beqnon
\langle\Phi^+_{p_1,\,\hj_1}(z_1,\bz_1,x_{1\,\alpha},\bx^{\alpha}_1)
\,\Phi^-_{p_2,\,\hj_2}(z_2,\bz_2,y_2^{\alpha},\by_{2\,\alpha})\rangle=
\delta(p_1-p_2)\,\delta(\hj_1+\hj_2)\,\prod_{\alpha=1}^{2}\frac{e^{-p_1(x_{1\,\alpha}x_2^{\alpha}+
\bar{x}^{\alpha}_1\bar{x}_{2\,\alpha})}}{|z_{12}|^{4h}}\, ,
 \eeqnon
where $h_{\pm} = \mp p \hj + p(1-p)$ and the $SU(2)_I$ symmetry is
manifest.

 By the same procedure the kinematical $x$ dependent part of
the three-point function between two `incoming' $p>0$ states and
one `outgoing' state with $p<0$ (by momentum conservation
$\delta(p_1+p_2-p_3)$), is given by
 \beqnon
K_{++-}(x_{1\,\alpha},x_{2\,\alpha},x_3^{\alpha};
\bar{x}^{\alpha}_1,\bar{x}^{\alpha}_2, \bar{x}_{3\,\alpha})
=\prod_{\alpha=1}^2\,
|e^{-x_3^{\alpha}(p_1x_{1\,\alpha}+p_2x_{2\,\alpha})}|^2
\,(x_{2\,\alpha}-x_{1\,\alpha})^{q_{\alpha}}\,(\bx^{\alpha}_2-
\bx^{\alpha}_1)^{q_{\alpha}}\, , \eeqnon
 where $L=- (\hj_1+\hj_2+\hj_3)=\sum_{\alpha}q_{\alpha}$ and $q_1, \, q_2 \in {\bf N}$.
 A sum over $q_1$ and $q_2$ is implicit. The important point we would like to stress
here is that in order to obtain an $SU(2)_I$ invariant expression
we should put together left and right moving parts, since the
relevant $SU(2)_I$ does not admit a chiral worldsheet description.

 Ward identities also fix the kinematical $x$ dependent part of the
extremal four-point function $\langle+++-\rangle$. The left piece
is
 \beqnon
K_{+++-}^{
{}^{LEFT}}(x_{1\,\alpha},x_{2\,\alpha},x_{3\,\alpha},x_4^{\alpha})=
\prod_{\alpha=1}^{2}\,
e^{-x_4^{\alpha}(p_1x_1+p_2x_2+p_3x_3)_{\alpha}}
(x_{3\,\alpha}-x_{1\,\alpha})^{q_{\alpha}}\, . \eeqnon
 Once again
we have an expression that is not $SU(2)_I$ chiral invariant. The
full correlator, including left and right movers, contains also
the dynamical part
 \beqnon
 \prod_{\alpha}^{2}\frac{1}{q_{\alpha}!}\,\left(\,C_{12}\,||f(z,x_{i\alpha})||^2+C_{34}\,||
 g(z,x_{i\alpha})||^2\,\right)^{q_{\alpha}}\, ,
 \eeqnon
 where
  $ f_{\alpha}(z,x_{i\alpha})=\sum_{i=1}^2 (x_{i\alpha}-x_{3\alpha}) f_{i}(z)$ and
  $\,g_{\alpha}(z,x_{i\alpha})=\sum_{i=1}^2 (x_{i\alpha}-x_{3\alpha})
 g_{i}(z)$;
 $f_i(z)$ and $g_i(z)$
 are given in terms of hypergeometric
 functions \cite{kda,cfs,bdkz1}. The coefficients $C_{12}$ and $C_{34}$ entering in this
 monodromy invariant combination are
 \beqnon
 C_{12}=\frac{\gamma(p_1+p_2)}{\gamma(p_1)\gamma(p_2)}, \qquad
 C_{34}=\frac{\gamma(p_4)}{\gamma(p_3)\gamma(p_1+p_2)}\, .
 \eeqnon
 Other correlators, in particular those with insertion of $p=0$ states, can be treated in
a similar though subtler way \cite{bdkz1}. As expected, string
amplitudes in the Hpp-wave can be obtained from those of
$AdS_3\times S^3$ by performing the Penrose limit on the $SL(2)$
and $SU(2)$ `charge variables', along the lines of
\cite{kda,bdkz1}. We thus see the dual role of the `charge'
variables: as a compact bookkeeping of the field content of a
given representation and as coordinate on a holographic screen
\cite{bdkz1,kirpiol}. In order to attempt any quantitative test of
this holographic interpretation one has to consider the
superstring.\\

In a quasi-free field realization, the bosonic generators of the
supersymmetric version of $\mathcal H_6$ algebra are \cite{hik}
 \begin{displaymath}
K=\oint\,{dz\over 2\pi i}\,\pd v(z), \qquad J=\oint\,{dz\over 2\pi
i}\,\pd\,u(z),
 \end{displaymath}
 \begin{displaymath}
P^+_{\alpha}=\oint \, {dz\over 2\pi i} \,e^{-i\,u}(i\pd
w_{\alpha}^*+\psi^+\psi^*_{\alpha})(z),\qquad P^{-\,\alpha}=\oint
\, dz\,e^{i\,u}(i\pd w^{\alpha}+\psi^-\psi^{\alpha})(z),
 \end{displaymath}
where the worldsheet fields contract according to
$$
u(z)v(w)\sim \log (z-w), \qquad w^*_{\alpha}(z)w^{\beta}(w)\sim
-\delta_{\alpha}^{\beta}\log(z-w)\, ,
$$
 \beqnon \psi^+(z)\psi^-(w)\sim
1/(z-w),\qquad \psi^*_{\alpha}\psi^{\beta}(w)\sim
\delta_{\alpha}^{\beta}/(z-w)\, .
 \eeqnon
 Depending on the choice of the internal manifold $\mathcal M$,
 super-$\mathcal H_6$ algebra will have 12 (for $T^4$) or
8 (for $K3$) fermionic generators in the left-moving sector (as
many in the right-moving sector). This can be seen decomposing the
left-moving spin field $S^{A}$ \cite{fms} belonging to the ${\bf
16}$ of $SO(9,1)$ into $({\bf 2}_L,{\bf 2}_L)_+ + ({\bf 2}_R,{\bf
2}_L)_- + ({\bf 2}_L,{\bf 2}_R)_- + ({\bf 2}_R,{\bf 2}_R)_+$ of
$SO(1,1)\times SO(4)\times SO(4)$. On the other hand BRST
invariance and GSO projection suggest the following form for the
four dynamical supercharges in the $({\bf 2}_R,{\bf 2}_L)_-$
 \beqnon
\mathcal{Q}_{dyn}^{-\dot\alpha a} = \oint \, {dz\over 2\pi i}
e^{-\phi/2} S^{- \dot\alpha a}\, . \eeqnon Similarly the four
kinematical supercharges in the $({\bf 2}_L,{\bf 2}_L)_+$  read
 \beqnon
\mathcal{Q}_{kin}^{+ \alpha a} = \oint \, {dz\over 2\pi i}
e^{-\phi/2} S^{+ \alpha a} e^{i \alpha u}\, .
 \eeqnon

These supersymmetries, common to both $T^4$ and $K3$, are in a
sense the same supersymmetries the superstring already had on
$AdS_3\times S^3\times \mathcal{M}$. However when taking the
Penrose limit of $AdS_3\times S^3\times T^4$, the BRST condition
relaxes and four extra supersymmetries in the $({\bf 2}_R,{\bf
2}_R)_+$ arise \nolinebreak\footnote{For $K3$ the number of
supersymmetries remain unchanged.}
 \beqnon \mathcal{Q}_{new}^{+\dot\alpha
\dot{a}}= \oint \, {dz\over 2\pi i} e^{-\phi/2}
S^{+\dot\alpha\dot{a}}\, .
 \eeqnon

In the BVW formalism only 8(+8) supersymmetries are manifest, in
agreement with the number of Green-Schwarz-like independent
spacetime odd variables. This works out right for $K3$
\cite{berkpp}. However, for $T^4$, the Penrose limit increases to
12(+12) the number of supersymmetries, making it impossible to
have all of them explicitly manifest in the hybrid formalism. It
seems that the 12(+12) manifest supersymmetries could only be
obtained by starting from the pure spinor formalism \cite{berk},
something that has not yet been done.\\

Once the explicit form of the supercharges is known, one can
determine the supermultiplet structure and identify BPS states.
The candidate vertex operators for $1/2$ BPS states read
\cite{hik}
 \beqnon
\mathcal{V}_{{}_{BPS}}^{(-),\alpha}=e^{-\phi}e^{i p v}\prod_i
\sigma_p^{(i)}:\psi^\alpha\Sigma_p^{(i)}:\quad ,
 \eeqnon
  where $\sigma_p^{(i)}$ and
$\Sigma_p^{(i)}$ are bosonic and fermionic twist fields. One can
easily check that half of the supercharges annihilate
$\mathcal{V}_{{}_{BPS}}^{(-),\alpha}$, while the remaining half
act on
it and build up a short supermultiplet.\\


As mentioned earlier, the boundary CFT dual to superstrings on
$AdS_3\times S^3\times \mathcal{M}$ is a non-linear $\sigma$-model
on the symmetric orbifold
$\mathit{Sym}^{N}(\mathcal{M})=(\mathcal{M})^{N}/{\mathcal
S}_{N}$, where $\mathcal S_{N}$ is the symmetric group of
$N=Q_1Q_5$ elements. Following the BMN conjecture \cite{bmn} that
relates light-cone momenta in the pp-wave background to conformal
dimensions and R-charges of the operators in the dual theory, the
spectrum of the super-CFT was shown to be given by
\cite{gms,gavnar} {\setlength\arraycolsep{2pt} \beqa
\mathrm{R\!-\!R}: \quad\quad \Delta-J&=&\sum_n N_n\,
\sqrt{1+\left(\frac{n\,g_s\,Q_5^{{}^R}}{J} \right)^2}
+g_s\,Q_5^{{}^R}\,\frac{L_0^{{}^\mathcal M}+\bar L_0^{{}^\mathcal
M}}{J}\, ,\nonumber\\
\mathrm{NS\!-\!NS}: \quad\quad \Delta-J&=&\sum_n N_n\,
\left(1+\frac{n\,Q_5^{{}^{NS}}}{J}
\right)+Q_5^{{}^{NS}}\,\frac{L_0^{{}^\mathcal M}+\bar
L_0^{{}^\mathcal M}}{J}\, .\nonumber \eeqa} \noindent where R-R
and NS-NS stand for
the nature of the 3-form flux.\\

  At a first scrutiny, it seemed
that the BMN correspondence failed to correctly  match the spectra
on the two sides even for states corresponding to operators with
large R-charge \cite{hik}. Nevertheless, it has been suggested
that, very much as for the original $AdS_3/CFT_2$ correspondence,
this might be due to the fact that the  `boundary CFT$_2$' is
sitting at the orbifold point, which is not the case for the bulk
description. In principle one can dispose of this mismatch by a
marginal deformation along the moduli space of the CFT$_2$
\cite{lunmat}. Alternatively one may, in principle, be able to
extrapolate the string spectrum to the symmetric orbifold point,
very much as in \cite{bms,beis,bbms} for the case
of $AdS_5/CFT_4$, and find precise agreement \cite{son}. \\

In addition one may envisage the possibility of computing
correlation functions in the `boundary theory', which results from
keeping only operators with large R-charge and where
superconformal symmetry has been traded for (or more technically
`contracted' into) the relevant supersymmetric version of ${\cal
H}_6$. Eventually they should be expressed in terms of the very
same charge variables that appear in the covariant string
amplitudes in the Hpp-wave. They play the role of coordinates on a
four-dimensional holographic screen, which is different from the
geometric boundary of the Hpp-wave \cite{berennast}, but captures
the essential features of the dynamics in the bulk \cite{bdkz2}.

Following \cite{kda,kirpiol}, we believe this is the way
holography works in the pp-wave limit of spaces with $AdS$
factors. Although we have only considered the simplest example of
the Hpp-wave resulting from a Penrose limit of $AdS_3\times
S^3\times \mathcal{M}$, we hope this may shed some light on the
fate of holography in the more realistic case of $AdS_5\times
S^5$. In particular using a manifestly covariant and
supersymmetric formalism, such the hybrid BVW \cite{bvw} or the
pure spinor formalism \cite{berk}, one should be able to address
the case with non-zero R-R
3-form flux \cite{berkpp} which represents the first hint to the case with R-R 5-form flux.\\
\\
\noindent {\bf Acknowledgements:}

It is a pleasure to thank the organizers of the `Tenth Marcel
Grossmann Meeting' for their invitation and CBPF-ICRA, especially
to Santiago P\'erez, for kind hospitality during my stay in Rio de
Janeiro. I'm grateful for discussions to N. Berkovits, N. Braga,
G. D'Appollonio, E. Kiritsis, K. Panigrahi, M. Prisco and A.
Sagnotti. I would like to thank M. Bianchi for collaboration and
for a critical reading of the manuscript. This work was supported
in part by INFN, the EC contracts HPRN-CT-2000-00122 and
HPRN-CT-2000-00148, the MIUR-COFIN contract 2003-023852 and the
NATO contract PST.CLG.978785. The author thanks the ICSC-World
Laboratory for financial support.

 {\setlength {\baselineskip}%
 {1\baselineskip}

}


\begin{thebibliography}{99}

\bibitem{malda} 
J.~M.~Maldacena,
Adv.\ Theor.\ Math.\ Phys.\  {\bf 2}, 231 (1998) [Int.\ J.\
Theor.\ Phys.\  {\bf 38}, 1113 (1999)] [arXiv:hep-th/9711200].

\bibitem{agmoo} 
O.~Aharony, S.~S.~Gubser, J.~M.~Maldacena, H.~Ooguri and Y.~Oz,
Phys.\ Rept.\  {\bf 323}, 183 (2000) [arXiv:hep-th/9905111].

\bibitem{gks} 
A.~Giveon, D.~Kutasov and N.~Seiberg,
Adv.\ Theor.\ Math.\ Phys.\ {\bf 2}, 733 (1998)
[arXiv:hep-th/9806194].

\bibitem{mo} 
J.~M.~Maldacena and H.~Ooguri,
J.\ Math.\ Phys.\  {\bf 42}, 2929 (2001) [arXiv:hep-th/0001053].


\bibitem{berk} 
N.~Berkovits,
JHEP {\bf 0004}, 018 (2000) [arXiv:hep-th/0001035].

\bibitem{bvw} 
N.~Berkovits, C.~Vafa and E.~Witten,
JHEP {\bf 9903}, 018 (1999) [arXiv:hep-th/9902098].

\bibitem{dolan} 
K.~Bobkov and L.~Dolan,
Phys.\ Lett.\ B {\bf 537}, 155 (2002) [arXiv:hep-th/0201027].

\bibitem{petropoulos} 
P.~M.~Petropoulos,
[arXiv:hep-th/9908189].


\bibitem{teschner}
J.~Teschner,
Nucl.\ Phys.\ B {\bf 546}, 390 (1999) [arXiv:hep-th/9712256].

\bibitem{hosom} 
K.~Hosomichi, K.~Okuyama and Y.~Satoh,
Nucl.\ Phys.\ B {\bf 598}, 451 (2001) [arXiv:hep-th/0009107].

\bibitem{bdkz2} M.~Bianchi, G.~D'Appollonio, E.~Kiritsis and
O.~Zapata, \emph{in preparation}.

\bibitem{seibwit} 
N.~Seiberg and E.~Witten,
JHEP {\bf 9904}, 017 (1999) [arXiv:hep-th/9903224].



\bibitem{penrose} R. Penrose, Differential Geometry and
Relativity, Reidel, Dordrecht 271 (1976).


\bibitem{blau} 
M.~Blau, J.~Figueroa-O'Farrill, C.~Hull and G.~Papadopoulos,
JHEP {\bf 0201}, 047 (2002) [arXiv:hep-th/0110242];
Class.\ Quant.\ Grav.\  {\bf 19}, L87 (2002)
[arXiv:hep-th/0201081].

\bibitem{sfetsos} 
K.~Sfetsos,
Phys.\ Lett.\ B {\bf 324}, 335 (1994) [arXiv:hep-th/9311010].

\bibitem{kda} 
G.~D'Appollonio and E.~Kiritsis,
Nucl.\ Phys.\ B {\bf 674}, 80 (2003) [arXiv:hep-th/0305081].


\bibitem{cfs} 
Y.~K.~Cheung, L.~Freidel and K.~Savvidy,
[arXiv:hep-th/0309005].



\bibitem{bdkz1} 
M.~Bianchi, G.~D'Appollonio, E.~Kiritsis and O.~Zapata,
[arXiv:hep-th/0402004].



\bibitem{bms} 
M.~Bianchi, J.~F.~Morales and H.~Samtleben,
JHEP {\bf 0307}, 062 (2003) [arXiv:hep-th/0305052].

\bibitem{beis} 
N.~Beisert,
Nucl.\ Phys.\ B {\bf 659}, 79 (2003) [arXiv:hep-th/0211032].


\bibitem{bbms} 
N.~Beisert, M.~Bianchi, J.~F.~Morales and H.~Samtleben,
[arXiv:hep-th/0310292].

\bibitem{son} 
J.~Son,
[arXiv:hep-th/0312017].

\bibitem{kirpiol} 
E.~Kiritsis and B.~Pioline,
JHEP {\bf 0208}, 048 (2002) [arXiv:hep-th/0204004].


\bibitem{hik} 
Y.~Hikida,
[arXiv:hep-th/0303222].

\bibitem{fms} 
D.~Friedan, E.~J.~Martinec and S.~H.~Shenker,
Nucl.\ Phys.\ B {\bf 271}, 93 (1986).


\bibitem{berkpp} 
N.~Berkovits,
JHEP {\bf 0204}, 037 (2002) [arXiv:hep-th/0203248].



\bibitem{bmn} 
D.~Berenstein, J.~M.~Maldacena and H.~Nastase,
JHEP {\bf 0204}, 013 (2002) [arXiv:hep-th/0202021].

\bibitem{gms}  
J.~Gomis, L.~Motl and A.~Strominger,
JHEP {\bf 0211}, 016 (2002) [arXiv:hep-th/0206166].

\bibitem{gavnar} 
E.~Gava and K.~S.~Narain,
JHEP {\bf 0212}, 023 (2002) [arXiv:hep-th/0208081].


\bibitem{lunmat} 
O.~Lunin and S.~D.~Mathur,
Nucl.\ Phys.\ B {\bf 642}, 91 (2002) [arXiv:hep-th/0206107].

\bibitem{berennast} 
D.~Berenstein and H.~Nastase,
[arXiv:hep-th/0205048].
\end{thebibliography}
\end{document}